\documentclass{emulateapj}

\slugcomment{}
\usepackage{graphicx}
\usepackage{longtable}

\shorttitle{A Very Large Structure at $z=3.78$}
\shortauthors{Lee et al. }
\begin{document}
\def\hh{\, h^{-1}}
\title{Discovery of a Very Large Structure at $z=3.78$}
\author{Kyoung-Soo Lee\altaffilmark{1,7}, Arjun Dey\altaffilmark{2,3},  Sungryong Hong\altaffilmark{2}, Naveen Reddy\altaffilmark{4,8}, \\Christian Wilson\altaffilmark{1}, Buell T. Jannuzi\altaffilmark{5}, Hanae Inami\altaffilmark{2}, Anthony H. Gonzalez\altaffilmark{6}}
\altaffiltext{1}{Department of Physics, Purdue University, 525 Northwestern Avenue, West Lafayette, IN 47907}
\altaffiltext{2}{National Optical Astronomy Observatory, Tucson, AZ 85726}
\altaffiltext{3}{Fellow, Radcliffe Institute for Advanced Study, Harvard University, 10 Garden Street, Cambridge, MA 02138}
\altaffiltext{4}{Department of Physics and Astronomy, University of California, Riverside, 900 University Avenue, Riverside, CA 92521}
\altaffiltext{5}{Steward Observatory, University of Arizona, Tucson, AZ 85721}
\altaffiltext{6}{Department of Astronomy, University of Florida, Gainesville, FL 32611}
\altaffiltext{7}{Visiting Astronomer, Kitt Peak National Observatory, National Optical Astronomy Observatory, which is operated by the Association of Universities for Research in Astronomy (AURA) under cooperative agreement with the National Science Foundation.}
\altaffiltext{8}{Sloan Research Fellow}

\begin{abstract}
We report the discovery of a large-scale structure containing multiple protoclusters at $z=3.78$ in the Bo\"otes field.  The spectroscopic discovery of five galaxies at $z=3.783\pm0.002$ lying within 1~Mpc of one another led us to undertake a deep narrow- and broad-band imaging survey of the surrounding field. Within a comoving volume of $72\times72\times25$~Mpc$^3$, we have identified 65 Lyman alpha emitter (LAE) candidates at $z=3.795\pm0.015$, and four additional galaxies at $z_{\rm{spec}}=3.730,3.753,3.780,3.835$.  The galaxy distribution within the field is highly non-uniform, exhibiting three large ($\approx 3-5\times$) overdensities separated by $8-14$~Mpc (physical) and possibly connected by filamentary structures traced by LAEs.  The observed number of LAEs in the entire field is nearly twice the average expected in field environments, based on estimates of the Ly$\alpha$ luminosity function at these redshifts.  We estimate that by $z=0$ the largest overdensity will grow into a cluster of mass $\approx10^{15}M_\odot$; the two smaller overdensities will grow into clusters of mass $(2-6)\times10^{14}M_\odot$. The highest concentration of galaxies is located at the southern end of the image, suggesting that the current imaging may not map the true extent of the large scale structure. Finding three large protocluster candidates within a single 0.3~deg$^2$ field is highly unusual; expectations from theory suggest that such alignments should occur less than 2\% of the time. Searching for and characterizing such structures and accurately measuring their volume space density can therefore place constraints on the theory of structure formation. Such regions can also serve as laboratories for the study of galaxy formation in dense environments.

\end{abstract}
  

\section{Introduction}

Hierarchical structure formation predicts that regions with the highest density are the sites of the earliest galaxy formation. The low-redshift descendants of such regions (virialized galaxy clusters) have been extensively studied and various relationships between galaxy properties and environmental density have been found \citep[e.g., morphology/star-formation-density relation;][]{dressler80,goto03}. However, the dominant physical processes that drive galaxy evolution in different environments and cause the observed present-day relationships remain poorly understood. Several studies have demonstrated that the red sequence that characterizes present-day clusters is in place as early as $z\sim2$ and that star-formation activity in these dense environments must have peaked at earlier epochs \citep[][]{blakeslee03,mei06,papovich10}. In addition, studies of a handful of clusters at redshifts $1<z<2$ reveal that star-formation is enhanced in dense regions: the slope of the star-formation--density relation observed at low-redshift appears to decrease and perhaps ``reverse'' at $z>1.5$, beyond which redshift cluster progenitors are expected to show enhanced star-formation relative to the field \citep[e.g.,][]{tran10,koyama13, brodwin13}. The evolution at even earlier epochs remains uncertain, primarily because of the lack of good protocluster candidates.

Identifying protocluster regions at early epochs  is important both to our understanding of large scale structure evolution and to the study of galaxy evolution in dense environments. However, 
finding un-virialized high-redshift ($z>2$) ``protoclusters'' is very challenging. At high redshift, identifying robust overdensities through purely photometric means is difficult because the commonly-used broad-band selection techniques result in broad redshift distributions (and many interloper populations) that significantly reduce the observed density contrast. Spectroscopy of large numbers of faint continuum-selected candidates requires expensive allocations of large telescope time. 
Unlike low-redshift ($z<1$) clusters, high-redshift protoclusters do not show detectable X-ray emission; at $z>2$ red sequences may not exist, or if they do, are difficult to identify observationally.
Despite these challenges, various heroic efforts have resulted in a heterogeneous sample of robust high-redshift protoclusters. The majority at $z>2$ have been found by targeting rare sources \citep[e.g., radio galaxies or Ly$\alpha$ blobs][]{miley04,palunas04,venemans02,venemans05,venemans07,steidel00,matsuda05,prescott08,mawatari12}; a handful of structures have been found in ``blank-field'' surveys \citep{steidel98,steidel05,shimasaku03,ouchi05}.  Recently, \citet{chiang14} identified a number of protocluster candidates at $1.8<z<3.1$ in the COSMOS field using photometric redshifts, but only two have been thus far confirmed using spectroscopic redshifts. The overall number of confirmed protoclusters (i.e., $\sim20$ at $z=2-6$) is still too low to allow for robust comparisons to the predictions of hierarchical structure formation models or for environmental studies of galaxy properties at different redshifts \citep{chiang13}.

In this work we report the discovery of an overdense region (roughly 0.3 deg$^2$ in size) containing multiple protocluster candidates lying within the Bo\"otes field of the NOAO Deep Wide-Field Survey \citep[NDWFS:][]{jannuzid99}. The presence of a protocluster was first signaled by the distribution of UV-continuum-selected galaxies. Here, we present the identification of robust Ly$\alpha$ line-emitting galaxy candidates in the field and use these to map and weigh the large-scale structure. In \S2, we describe the initial search technique that resulted in identifying a protocluster candidate and thus selecting the field for further study. In \S3, we describe new data that we obtained to map the distribution of Ly$\alpha$ emitters in the field; we detail the selection of the candidates and assess their robustness. In \S4 we describe the distribution of galaxies in the structure and assess the overdensities of the protocluster candidates in the field. We conclude in \S5.

Throughout, we use  the WMAP7 cosmology $(\Omega, \Omega_\Lambda, \sigma_8, h)=(0.27, 0.73, 0.8, 0.7)$ \citep{wmap7}; at $z=3.78$, the angular scale is 7.347kpc/\arcsec (physical). Distance scales are presented in units of comoving Mpc unless noted otherwise. All magnitudes are given in the AB system \citep{oke83}. 

\section{Observations and Sample Definition}

\subsection{Identifying a Field with a Protocluster Candidate}

We have been undertaking a search for UV-luminous star-forming galaxies at $z\approx 3.7\pm 0.4$ in the Bo\"otes field of the NDWFS. We refer the reader to \citet{lee11} and \citet{lee13}  for details regarding the search, but briefly describe the procedure here. We applied a standard Lyman-break galaxy color-selection technique to the NDWFS $B_WRI$ data and selected candidates based on the following criteria (same as Lee et al.~2011, but in AB mag):
\begin{eqnarray}
(B_W-R)> 3~(R-I)+1.74~ \cap ~(B_W-R)\geq 1.8~~~~~~~~~\nonumber \\
 (R-I) \geq -0.54~~  \cap ~~ S/N(R) \geq 3 ~~ \cap  ~~~S/N(I)\geq 7 ~~~~~~~~~
\end{eqnarray}

For sources undetected in the $B_W$-band, we used the $2\sigma$ limit. A subset of the candidates was observed spectroscopically using the DEIMOS instrument at the Keck II telescope of the W. M. Keck Observatory; these observations confirmed the redshift distribution of the sample and characterized the selection function \citep{lee11}. These spectroscopic observations also resulted in the discovery of five galaxies lying at redshifts $3.782 < z < 3.787$, within 1~Mpc (projected physical distance) from one another \citep[see Figures 11 \& 12 and \S5.1 in][]{lee13}. We interpreted this alignment of galaxies as signaling the presence of a coherent large scale structure and set out to measure its extent and spatial overdensity using Ly$\alpha$ emitting galaxies as a tracer population. 

\subsection{New Observations}

In May and June 2012, we obtained deep imaging of the candidate protocluster field using the Mosaic 1.1 Camera on the Mayall 4-m telescope of the Kitt Peak National Observatory. Images were obtained through both broad-band ($B_WRI$) and narrow-band (WRC4) filters centered on the field NDWFSJ1426+3236\footnote{For details regarding the NDWFS, see http://www.noao.edu/noao/noaodeep/. Field coordinates are at http://www.noao.edu/noao/noaodeep/ndwfs\_op\_new.html}. The WRC4 filter (designed to sample \ion{C}{4} emission in Wolf-Rayet stars; KPNO filter \# k1024) has a central wavelength (in the KPNO 4m f/3.1 prime-focus corrector beam) of 5819\AA\ and a full-width at half maximum of 42\AA; it is therefore well placed to sample Ly$\alpha$ emission at $3.769<z<3.804$ (i.e., 25 comoving Mpc).  We used individual exposure times of [20,20,10,10] minutes for the [$B_W$,WRC4,$R$,$I$] bands respectively and dithered the telescope between exposures by $<2$~arcmin. After discarding frames taken with delivered image quality $>1.1$\arcsec, the effective total exposure times of the new imaging are 8.3, 3.0, 5.2, and 0.8 hours for the WRC4, $B_W$, $R$, and $I$ bands respectively. For each frame, we updated the astrometry (employing the {\tt IRAF}\footnote{IRAF is distributed by the National Optical Astronomy Observatory, which is operated by the Association of Universities for Research in Astronomy (AURA) under cooperative agreement with the National Science Foundation.} task {\tt msccmatch}) using stars identified in the Sloan Digital Sky Survey DR7 Catalog, and reprojected it to a common tangent point with a pixel scale of 0.258\arcsec/pixel. The same procedure was repeated for the NDWFS frames after recomputing the MOSAIC camera image distortions using a large catalog of SDSS DR7 sources. 

The $B_W$, $R$, and $I$ band images are photometrically tied to the existing NDWFS images. The relative intensity scale of each reprojected frame was determined with respect to the NDWFS data using the {\tt mscimatch} task in the IRAF {\tt mscred} package. All frames of a given band were combined to construct the final image stack using a weighted average, with the relative weight inversely proportional to the variance of sky noise measured in the reprojected frames. 
We masked regions near the field borders with less than 20\% of the maximum exposure time and areas near bright saturated stars and their diffraction spikes.  The effective area of the final mosaic is 0.317 deg$^2$.

The photometric calibration for the $B_W,R,I$ images is referenced to the NDWFS images.  For the WRC4 band, we determined the photometric zero point using spectrophotometric standard stars observed during the run. Combining the new data with the existing NDWFS data results in effective exposure times of 8.3, 5.3, 6.9, and 4.1 hours for the WRC4, $B_W$, $R$, and $I$ bands respectively, resulting in $5\sigma$ limiting magnitudes of 24.91, 26.96, 26.34, and 25.55 AB mag (measured on blank sky regions within 2$\arcsec$ diameter circular apertures). The delivered image quality is $0.9\arcsec-1.0\arcsec$ in the final stacks. 

We estimate the detection completeness by inserting into the stacked images artificial sources with similar sizes and morphologies as real high-redshift galaxies. We find that the 90\%, 50\%, and 10\% completeness limits for source detection are 23.82, 24.22, 24.64 mag, respectively, for the WRC4 band, and 24.07, 24.78, 25.26 mag  for the $I$- band. 

\subsection{Selection of $z\approx3.78$ LAE Candidates}\label{LAE_selection}

Our primary goal is to investigate the spatial extent of the presumed galaxy overdensity at $z=3.78$ as traced by Ly$\alpha$-emitting galaxies (LAEs).  We identify candidate LAEs as sources that show excess emission in the WRC4 band, i.e., unusually blue narrow-band-to-broad-band colors. 

\begin{figure*}
\epsscale{1}
\plotone{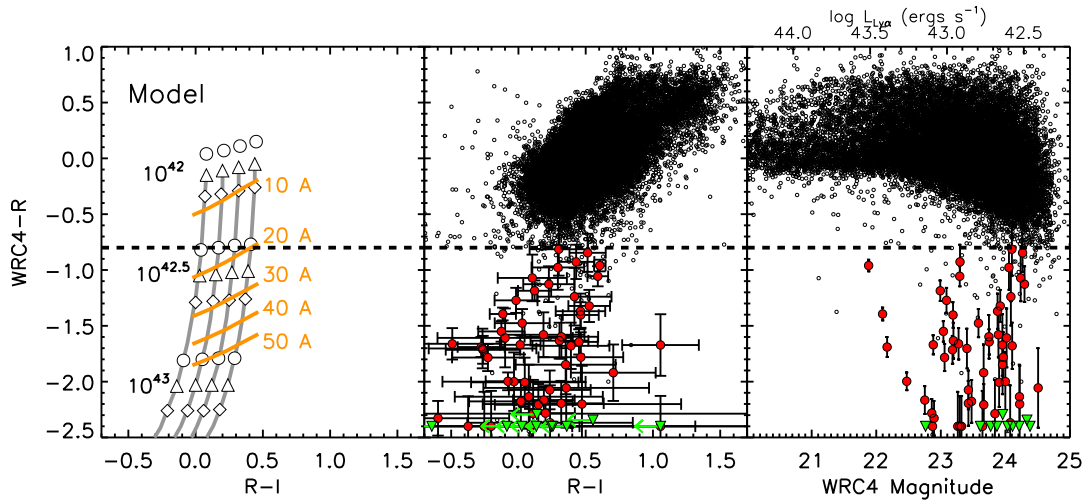}
\caption[selection]{
Left: Four grey lines show the color evolution with increasing Ly$\alpha$ luminosities (from top to bottom) at four different reddening values. 
Triangles (diamonds, circles) show the Ly$\alpha$ luminosities, $10^{42.0}$, $10^{42.5}$, $10^{43.0}$, and $10^{43.5}~{\rm ergs~s}^{-1}$ at the continuum $R$-band magnitude of 25.5 (25.8, 25.2), respectively (see \S\ref{overdensity_estimate} for detail). Orange lines represent the Ly$\alpha$ rest-frame equivalent widths of (from top to bottom) 10, 20, 30, 40, and 50\AA\ at different continuum $R-I$ colors. The ${\rm WRC4}-R$ color cut is chosen to match (approximately) the expected color of $W_0=20$\AA\ sources at the median $R-I$ color.; 
Middle:  $WRC4-R$  vs. $R-I$ colors are shown for all sources detected in the WRC4 band.  Photometric LAE candidates selected according to equation 2 (i.e., below the thick dashed line) are shown as circles ($S/N(R)\geq 2$) or triangles ($S/N(R)<2$). The sources that are formally undetected in the $I$-band are shown with upper limits. Galaxies with $({\rm WRC4}-R)\leq -2.4$ are shown at the color position of $-2.4$. 
Right:  $({\rm WRC4}-R)$ colors as a function of WRC4-band magnitude. The upper limits for the sources with ${\rm S/N} (R, I) < 2$ (i.e., undetected in $R$ or $I$) are shown as triangles. The Ly$\alpha$ luminosities corresponding to the WRC4 magnitude are indicated on top at the $R$-band magnitude fixed to the median value of  25.5. 
  }
\label{lae_selection}
\end{figure*}

We identify LAE candidates as follows. We created source catalogs in the four bands by running the SExtractor software \citep{bertina96} in ``dual image mode'' on all the image stacks. We used the WRC4-band image as the detection image in order to select LAEs that show detectable Ly$\alpha$ emission, but which may be undetected in the continuum broad-band filters. The SExtractor parameter MAG\_AUTO is used to estimate the total magnitude, while colors are computed from fluxes within a fixed isophotal area (i.e., FLUX\_ISO). LAE candidates at $z=3.78$ are selected to satisfy:
\begin{eqnarray}
({\rm WRC4}-R) <-0.8 ~~ \cap  ~~~S/N({\rm WRC4})\geq 7 ~~~~~~~~~ \nonumber \\
\cap ~~ [(B_W-R)>1.8 ~~\ \ \cup~~ S/N(B_W) < 2] ~~~~~~~~~~~~~~
\end{eqnarray}
where S/N is computed within the isophotal area. 
Red $B_W-R$ colors ensure that there is a strong spectral break (due to the Lyman break and Ly$\alpha$ forest) in the $B_W$ band such that candidates lie at $z>3.2$. 
Our LAE photometric  selection is illustrated in the left panel of Figure~\ref{lae_selection}. 

In order to understand the sample selection criteria, we synthesized the ($B_W-R$) and (${\rm WRC4}-R$) colors of model LAEs with a large range in rest-frame UV continuum slope, Ly$\alpha$ emission line equivalent width, and Ly$\alpha$ luminosity. 
First, we created a model LAE spectrum as follows. We used the stellar population synthesis model of \citet{bc03} to compute the continuum spectrum of a galaxy with a constant star formation history observed at the population age of 100 Myr, assuming a Salpeter initial mass function and solar metallicity. We accounted for attennuation by intergalactic hydrogen using the model of \citet{madau95}. We adopted the dust extinction model of \citet{calzetti00} and varied the reddening parameter $E(B-V)$ until the predicted $R-I$ color approximately matched the observed value $\langle R-I \rangle \sim 0.2$ as shown in Figure \ref{lae_selection} (left panel). To the continuum spectrum, we added\footnote{In practice, the line flux was computed separately from continuum flux due to coarse resolution of the \citet{bc03} galaxy templates.} a Ly$\alpha$ emission line with a Gaussian line profile centered at 1215.67~\AA\ and an intrinsic line width of 3~\AA. Exact values assumed for the line width and reddening parameter are not important as long as they reasonably reproduce the observed galaxy colors and line FWHM. 

Model spectra were generated for a range of Ly$\alpha$ rest-frame equivalent widths, $W_0$, and luminosities and used to predict ${\rm WRC4}-R$ colors (see Figure \ref{lae_selection}, left panel). The color criterion ${\rm WRC4}-R \leq -0.8$ corresponds to $W_0\gtrsim20$~\AA\ at $z=3.785$. Figure \ref{lae_selection} also shows the expected $({\rm WRC4}-R)$ and $(R-I)$ colors at different extinction/reddening values. At a fixed reddening value, large open triangles (diamonds, circles) show the  Ly$\alpha$ luminosities of, from top to bottom, $10^{42.0}$, $10^{42.5}$, $10^{43.0}$, and $10^{43.5}~{\rm ergs~s}^{-1}$ at the continuum $R$-band magnitude of 25.5 (25.2, 25.8)  mag. The median $R$-band magnitude is 25.5, with the majority at $R=25.5\pm$0.6  mag. The limiting Ly$\alpha$ luminosity of our survey is   $\sim10^{42.6}~ {\rm ergs~s}^{-1}$. 

We find 65 LAE candidates within the total survey area of 0.317~deg$^2$, corresponding to an average surface density of $0.057\pm0.007$~arcmin$^{-2}$.  
Ten of the LAE candidates are only detected in the WRC4 band and have no $R$ or $I$-band continuum detections (shown as triangles in Figure~\ref{lae_selection}. Approximate Ly$\alpha$ luminosities corresponding to their WRC4-band magnitudes are indicated on the top abscissa; the majority of our sources have $L_{\rm{Ly}\alpha}\gtrsim 10^{42.5}~{\rm ergs~ s}^{-1}$ and in the WRC4 magnitude range of $23.0-24.5$, while there are six LAEs brighter than ${\rm WRC4}\leq 22.8$ ($L_{{\rm Ly}\alpha}\sim(1.5-3.0)\times 10^{43}~{\rm ergs~ s}^{-1}$). 

\subsection{Selection of $z\sim3.7$ Star-Forming Galaxy Candidates}\label{LBG_selection}

In addition to the LAE candidates, we also used the deeper $B_WRI$ broad-band imaging to identify a sample of Lyman-break star-forming galaxies (LBGs) at $z\sim3.7\pm0.4$ using the selection criteria described in equation (1).  The reliability of $z\sim3.7$ LBG candidates selected according to these criteria has been investigated by \citet{lee13}.

We identified 428 LBG candidates within our field, i.e., an observed candidate surface density of 0.375$\pm$0.018~arcmin$^{-2}$. Accounting for the possible contamination by interlopers \citep[measured by][to be $11-28$\%]{lee13}, this suggests an LBG surface density in the field of 0.30$\pm$0.05~arcmin$^{-2}$, comparable to that reported in \citet{bouwens07}. The broad-band colors of the LBG candidates are shown in Figure~\ref{lbg_selection} as black dots within selection window. Their angular distribution is shown in Figure~\ref{distribution} (grey circles). 
 
\subsection{Reliability of the LAE Candidates}

Although the bulk of our sample awaits spectroscopic confirmation, four of the five spectroscopically confirmed UV-continuum-selected Lyman Break galaxies are members of our LAE candidate list.

\begin{figure}[t]
\epsscale{1.2}
\plotone{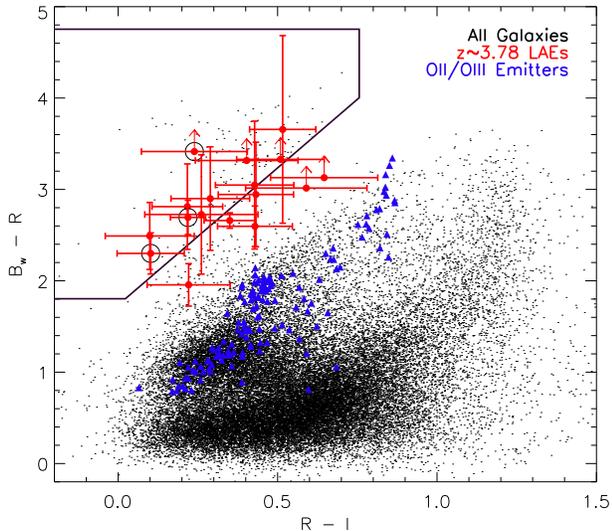}
\caption[selection]{
The $B_W-R$ vs $R-I$ colors of all $I$-band detected sources are shown in dots, together with the dropout criteria outlined in solid line. Small open circles (triangles) represent galaxies satisfying the Lyman break galaxy (LBG) selection with (without) secure $B_W$ detection. The colors of the $I$-band detected LAEs are indicated in red symbols. Those accompanied by larger open circle represent the LAEs among the six brightest NB sources (${\rm WRC4}<22.8$). More than two thirds of the line-emitting candidates are not detected in the $I$-band image. The majority of the $I$-band detected LAEs have colors similar to LBGs with the exception of one. Lower-redshift sources which would have been observed with substantial narrow-band excess due to their [O~{\sc iii}] or [O~{\sc ii}] emission are shown as blue triangles. 
  }
\label{lbg_selection}
\end{figure}

\begin{figure*}
\epsscale{1.0}
\plottwo{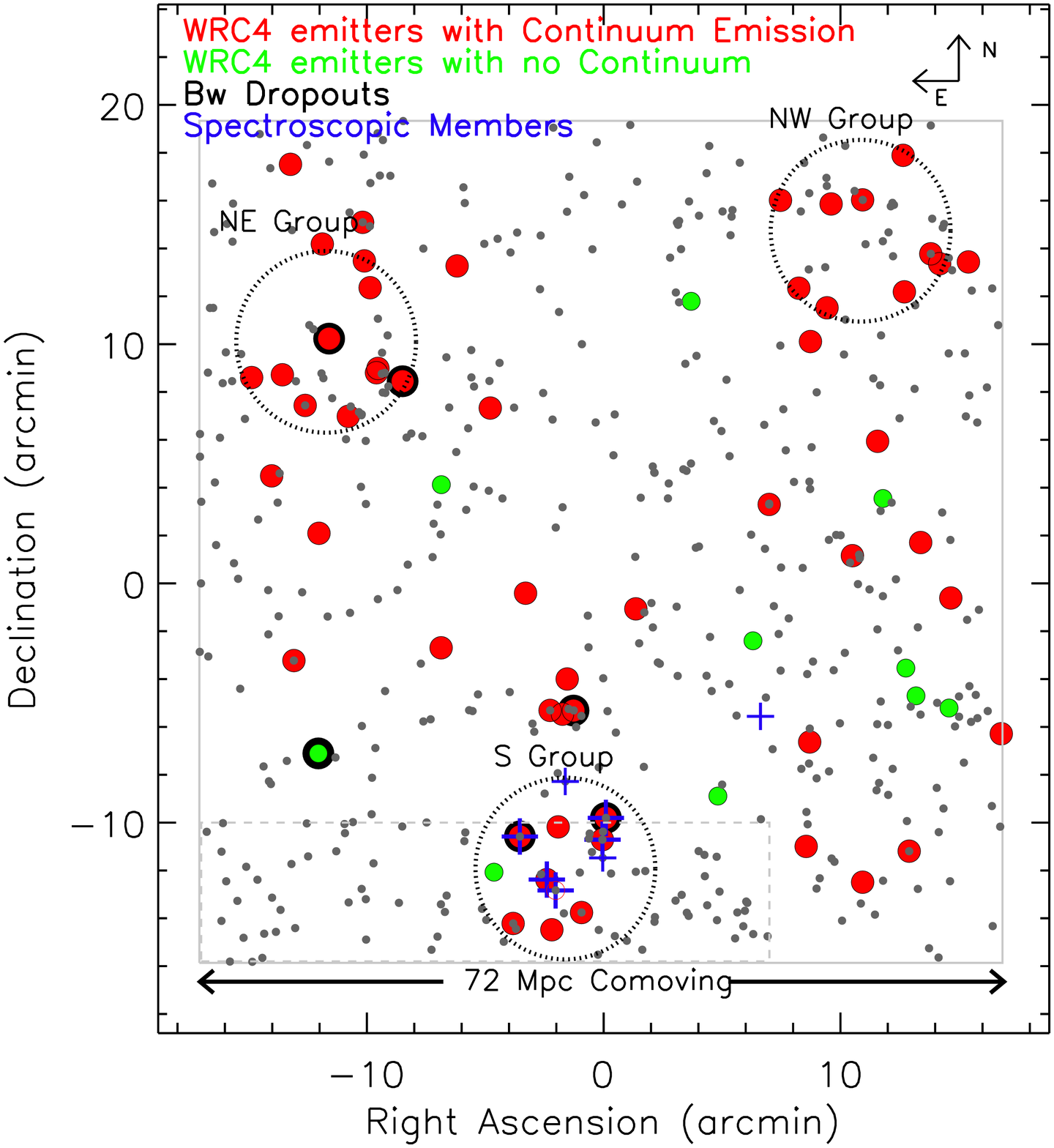}{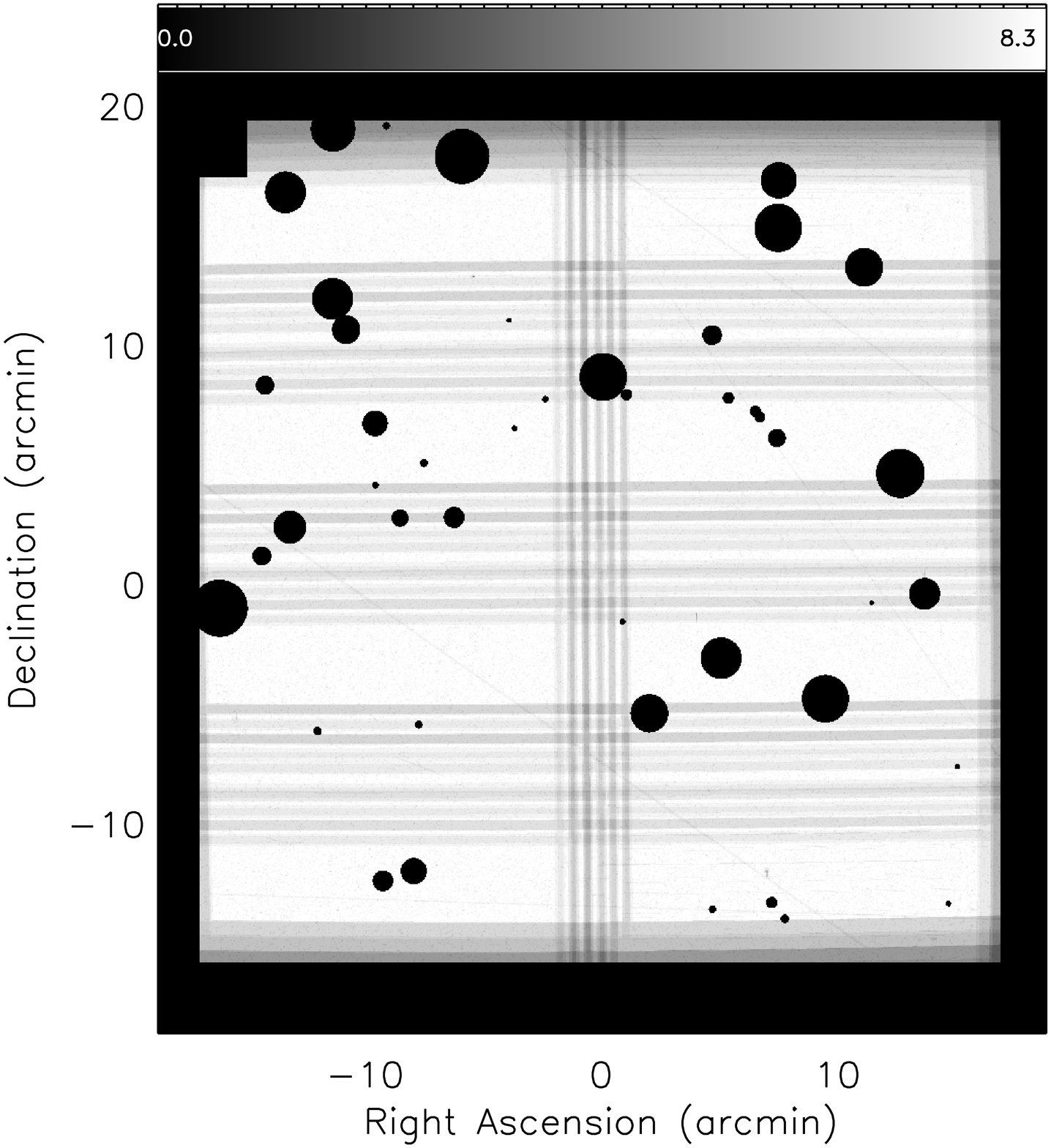}
\caption[distribution]{
Left: The angular distribution of the LAEs with/without continuum emission (red/green circles) and $z\sim3.7$ LBGs (small filled circles) is illustrated together with the positions of known spectroscopic sources at $z=3.730-3.835$ (i.e., within 50~Mpc from the center of the S group; crosses). The field boundaries are indicated with a large grey box. Six brightest WRC4 sources are marked by  large open circles. Three densest regions, dubbed the NE, NW, and S groups, are marked by dotted 8~Mpc radius circles. The brightest WRC4 sources (i.e., the most luminous LAEs) are within or near these groups. 
Right: The exposure map and object mask for the field. Objects falling within these masks were removed from our candidate lists. The inhomogeneous galaxy distribution is not the result of masking or inhomogeneous depth across the field. The color bar (top) indicates the effective exposure time in the WRC4 band in units of hours. 
  }
\label{distribution}
\end{figure*}

Photometrically-selected LAE samples can be contaminated by low-redshift interloping galaxies where, e.g., emission lines of  [O~{\sc iii}] and [O~{\sc ii}] fall within the bandpass of the narrow-band filter.  Low-redshift interlopers at $0.158<z<0.166$ with strong [OIII]$\lambda$5008 emission, or at $0.553<z<0.569$ with strong [OII]$\lambda$3729 emission, will exhibit a $(WRC4-R)$ color excess and can potentially contaminate the LAE candidate sample. Although they may satisfy the $(WRC4-R)$ criterion, they may be discriminated by the $(B_W-R)$ selection criteria. Figure~\ref{lbg_selection} shows the $(B_W-R)$ and $(R-I)$ color-color diagram of all detected sources in the field (dots); the subset of 17 LAE candidates detected in both $R$ and $I$ bands are also shown (red filled circles). Also shown are spectroscopically confirmed galaxies from the AGN and Galaxy Evolution Survey \citep[AGES;][]{kochanek12} that lie within the redshift ranges $0.158<z<0.166$ and $0.553<z<0.569$ (blue triangles). Although the AGES survey targeted bright galaxies ($I<20$~Vega mag), none of them have colors that overlap with the LAE sample. Two of the 17 sources lie between the diagonal LBG selection criterion boundary and the low-redshift galaxy sequence, but there is no overlap between the populations. Accounting for upper limits, we find that only three of the 55 LAEs with continuum detection lie within 0.2~mag of the low-redshift galaxy sequence. We therefore conclude that the contamination is no more than 6\%.

While LAEs are typically powered by star-formation, some of the more luminous emitters could be AGN \citep[e.g.,][]{ouchi08}. While this is not an issue for tracing the structures, we briefly consider whether there is evidence for AGN within our LAE candidate sample. The X-ray data that exist in the field \citep[from the Chandra XBo\"otes Survey;][]{kenter05} are too shallow to place firm constraints on AGN contamination for the bulk of the sample. However, none of of the LAE candidates  are detected in the X-ray.  All but two LAE candidates are covered by existing {\it Spitzer} MIPS 24$\mu$m imaging. Among them, only one is clearly detected. Visual inspection of the morphologies show that most of them are spatially extended ($\rm{FWHM}\geq 4.5~\rm{pixel}$ or 1.2\arcsec) and not dominated by a point source component that may suggest the presence of an AGN. Two have existing rest-frame UV spectra \citep{lee13}, but neither shows C~{\sc iv} or N~{\sc v} emission indicative of AGN.  We therefore conclude that the contamination of the LAE candidate population by luminous AGN is not significant.

Some guidance may also be found from other work. \citet{ouchi08}, whose survey is the closest match to ours in terms of redshift and imaging depth, did not find any clear evidence of contamination in their $z=3.7$ LAE spectroscopic sample. After similar narrow-band searches of LAEs around high-redshift radio galaxies and their spectroscopic followup, \citet{venemans07} reported that contamination rate averaged over all fields is $\sim 9$\% (in some fields, they found zero contamination).  

While we cannot be certain without spectroscopy of the candidates, we conclude that the selection of the LAE candidates is robust and that most, if not all, lie at $z\approx3.78$. In what follows, we use the spatial distribution of LAEs to estimate overdensities; any low-$z$ interloper contamination, if randomly distributed, will result in an underestimate of the true overdensity.
 
\section{Results}\label{results}
\subsection{Distribution of LAEs and LBG candidates}\label{galaxy_distribution}

In this section, we discuss the angular distribution of the galaxies within the field. We consider two samples separately: (1) the 65 LAE candidates supplemented by the spectroscopically confirmed galaxies (which adds one extra galaxy to this subset); and (2) the 428 LBG galaxy candidates. The distributions of these two subsets are shown in the left panel of Figure \ref{distribution}. 

The LAE distribution within the field (which covers a volume of $72\times72\times25$~Mpc at $z=3.785$; Figure \ref{distribution}) is highly inhomogeneous. In particular, there are three regions (denoted as ``NE'', ``NW'', and ``S'' groups) each containing $9-10$ LAEs within a 8~Mpc (or 3.8\arcmin) radius circle. Given the observed average LAE surface density in this field of 0.057~arcmin$^{-2}$ ($=65/(0.317\times 60^2)$), the expected number of LAEs within a 3.8\arcmin\ radius circle is $\sim 2.6\pm0.3$. Hence, these three regions have surface densities that are factors of $3.5-3.8$ higher than the average in this field, or surface overdensities of $\delta_\Sigma\equiv(\Sigma-\bar\Sigma)/\bar\Sigma)\approx2.5-2.8$.  Since the average surface density measured in this field is likely biased by the presence of the structures (see \S4.2), these are underestimates of the true surface overdensities.

The three groups lie $\approx40-60$~comoving~Mpc (i.e., $8.4-12.5$~Mpc physical) apart from one another (in projection), and appear to be connected by filaments traced by the LAE distribution. The three spectroscopically confirmed galaxies at $z\ne3.78$ (shown as blue crosses without red circles) near the S group may trace a filament running along line of sight. There are also regions  devoid of any LAE candidates (in the northern and central regions and the southeast corner). These areas have comparable photometric depth to the regions in which galaxy overdensities are found, and are not significantly affected by the object mask (see right panel of Figure~\ref{distribution}). Hence, the observed inhomogeneity of the galaxy distribution is likely real. Five of the six highest line-luminosity sources discussed in \S\ref{LAE_selection} (the large black circles in Figure~\ref{distribution}, left) lie within or near the three overdense groups: two in the NE group; two in the S group; and one just north of the S group, where there is a concentration of four LAEs. Only one of the six resides far from the galaxy overdensities northeast of the S group. 

The distribution of LBG candidates also appears highly inhomogeneous. A large number of LBGs are observed at the southern edge of the field near (but not coincident with) the S group. Within the 24\arcmin$\times$6\arcmin\ area (dashed grey line in Figure~\ref{distribution} left), there are 83 LBGs, 50\% larger in number than that expected from the observed field average (i.e., $0.375~\rm{arcmin}^{-2} \times (6\times 24)~\rm{arcmin}^2 =54$; see \S~\ref{LBG_selection}). 

\begin{figure*}
\epsscale{1.}
\plottwo{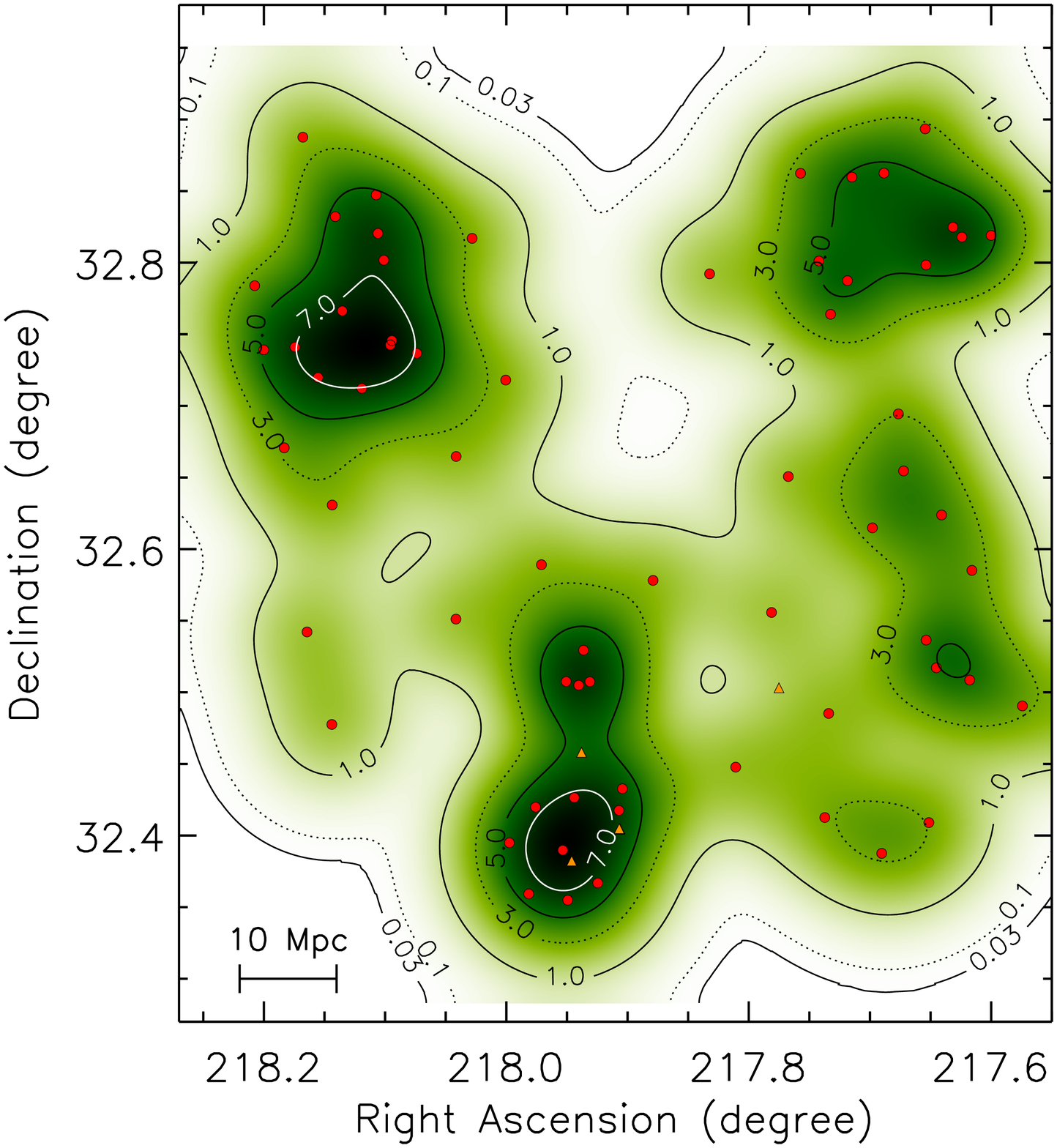}{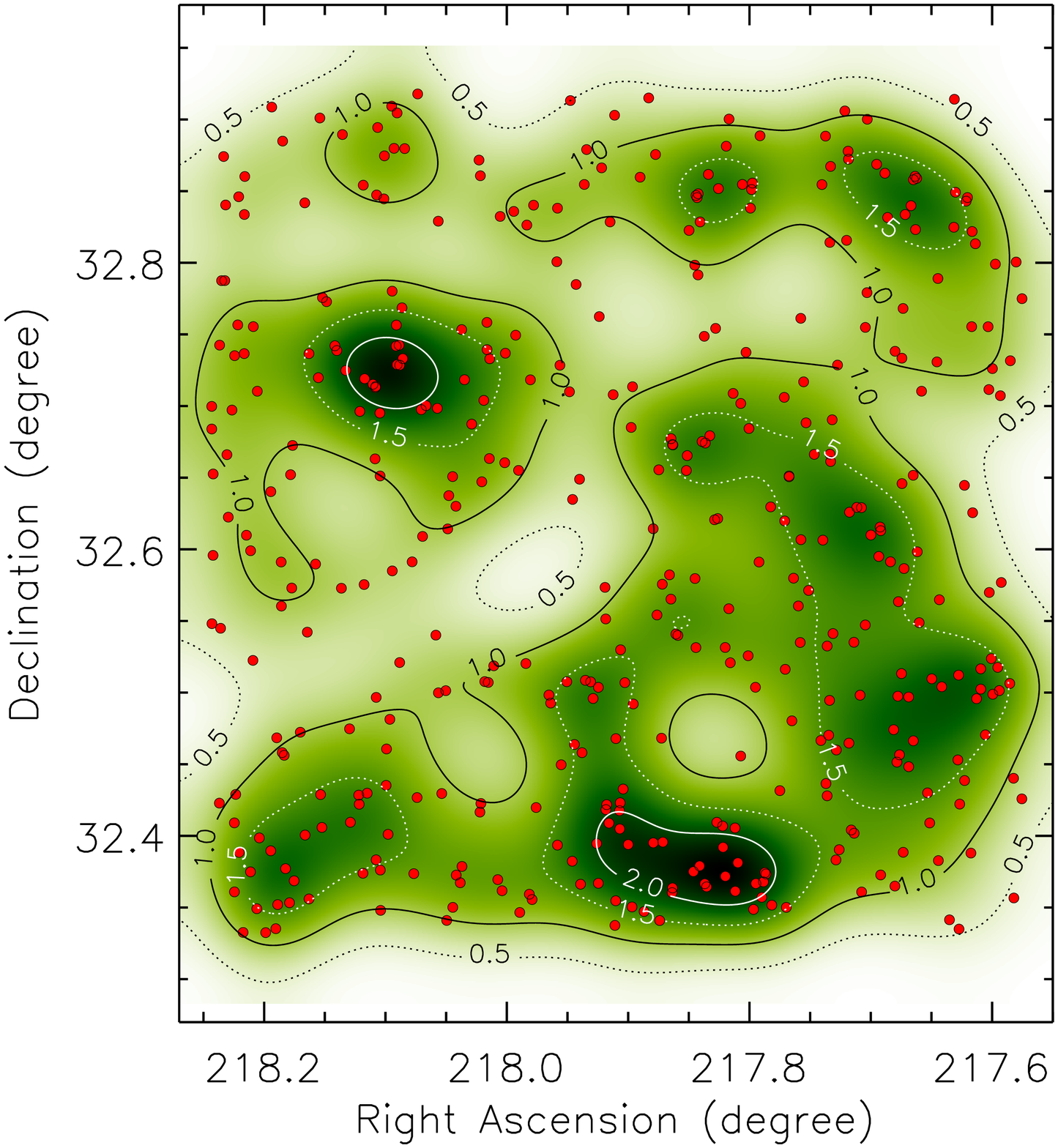}
\caption[structure_map]{
Local density fluctuation of the $z=3.78$ LAEs (left) and $z\sim3.7\pm0.4$ LBGs (right): the structure is constructed by smoothing the positions of the member galaxies with a gaussian kernel (FWHM=10~Mpc). The contours show the relative surface density with respect to the local density. The locations of the three high-overdensity regions are clearly visible in the LAE map, each observed with $\geq5\times$ the field density at the core.  Interestingly, the regions with the highest LAE density are also observed with the highest density of LBGs.   
}
\label{structure_map}
\end{figure*}

To better visualize and quantify the angular distribution of LAEs and LBGs, we create two-dimensional density maps of the field. We first apply a gaussian kernel of FWHM=10~Mpc (i.e., characteristic radius of 4.25~Mpc) independently to the positional map of the 65 LAE candidates and that of the 428 LBG candidates (see Figure~\ref{structure_map}). The contours show the relative local density with respect to the field. The discussion of the LAE field density estimate is given in \S\ref{overdensity_estimate}. For the LBGs, we assume the observed LBG surface density as the field average. 

The locations of the three high-overdensity regions, discussed in \S\ref{galaxy_distribution} and shown in Figure \ref{distribution}, are apparent in the LAE density map. Each group resides in a region with the surface density at the core $\geq 5\times$ the field average. The S group is the most significant overdensity, containing 9 LAEs and 3 additional spectroscopic members. A smaller, compact group of LAEs lies $\sim$6\arcmin\ ($\sim$12.6 comoving Mpc; 2.6 physical Mpc) to the north. We varied the smoothing scales to test the robustness of the overdensity regions. While the overall morphology of the LAE density map changes slightly depending on the smoothing scale, all three groups are recovered with smoothing scales between $8-16$~Mpc. 

Despite the broader redshift distribution resulting from the LBG selection criteria, the LBG density map shows a striking resemblance to that derived for the LAEs. The two regions of highest LBG density are located in or near the S and NE LAE groups.  The LAE and LBG distributions associated with the S group are slightly offset from each other, with the LBG surface density forming an elongated structure extending nearly 3~Mpc (physical) away from the peak of the LAE surface density. The NW group of LAEs also resides in a region where the LBG density is $\sim1.5\times$ that of the ``field''.  The density contour around the S group of LBGs  extends northward, mirroring the distribution of LAEs. The morphology of LBG density in the region between the NW and S group is also similar to that observed in the LAE map. These similarities lend support to the idea that these are coherent structures in the galaxy distribution. 

One notable difference between the LBG and LAE maps is the presence of a large population of LBGs in the southeast corner of the field ($1.5\times$ the field); no LAEs are observed there. Differences of this nature might suggest that these are regions where the LBG redshifts (and the dominant structure) do not fall within the LAE redshift selection window, but are either foreground or background structures.

The line-of-sight distances sampled by LBG and LAE candidates are vastly different: the front-to-back comoving distance of the LAE sample is $\sim25$~Mpc, just $4-5$\%  of that for the LBG sample, i.e., 480 (640)~Mpc, assuming $\sigma_z=0.3~(0.4)$ around $z=3.7$ \citep{lee13}.  Nevertheless, the spatial density distributions of the two candidate populations resemble each other, and do not appear to be affected by any spatially varying selection criteria. The close resemblance of the two distribution lends strong support to the fact that many of the LBG candidates must be part of the large-scale structure traced by the LAEs, and that the discovered structure genuinely represents highly significant matter density peaks traced by both LAEs and LBGs. In the section that follows, we attempt to quantify the overdensities and predict their present-day descendants.

\subsection{Galaxy Overdensity in the Protocluster Field}\label{overdensity_estimate}
In hierarchical galaxy formation theory, cluster formation takes place in the most highly overdense regions.  Well before their final assembly and subsequent virialization, protocluster regions can be identified via their overall enhancement in space and/or surface density of galaxies \citep[e.g.,][]{chiang13, chiang14}. Predicting the total mass overdensity, $\delta_m$, from the observed galaxy overdensity, $\delta_{\rm{gal}}$ is uncertain without full knowledge of the star- and galaxy-formation physics within dark matter halos; nevertheless, the observed galaxy overdensity can be used (in a relative sense) to gauge the significance of a given structure in the cosmological context and to predict the final mass of its descendant in the present-day universe \citep{chiang13}. Here, we quantify the significance of the observed galaxy overdensity and begin by estimating the expected number of LAEs in field environments. 

We estimate the expected number of LAEs selected by our survey by integrating the Ly$\alpha$ luminosity function derived by \citet{ouchi08} for a $z\approx3.7$ LAE sample. Assuming a Schechter function form,
$\phi(M)=(0.4 \ln 10) \phi^* ~10^{0.4(M_{\rm{Ly}\alpha}^*-M)(1+\alpha)} \exp[-10^{0.4(M_{\rm{Ly}\alpha}^*-M)}) ]$,
we used $\phi^*=(3.4^{+1.0}_{-0.9}\times10^{-4})~{\rm Mpc}^{-3}$, $L^*_{{\rm Ly}\alpha}=(10.2^{+1.8}_{-1.5}\times10^{42})~{\rm ergs~s}^{-1}$, and $\alpha=-1.5$, weighted by the filter throughput and the completeness function which defines the depth of our imaging data.  
The expected number of LAEs in a given WRC4 magnitude bin ($m_k$) and redshift bin ($z_j$) is given by:
\begin{equation}
N_{\rm{LAE}}(z_j,m_k)=V_j~p(m_k)~\mathcal{T}(z_j)~\int_{L_k} \phi(L)~dL
\end{equation}
where $\phi(L)$ is the Ly$\alpha$ luminosity function; $\mathcal{T}(z)$ is the redshift selection function, defined from the effective filter transmission (which includes the atmosphere, telescope, and instrument), $T(\lambda)$, as $\mathcal{T}(z_j) \equiv T_{WRC4}[\lambda=1215.67\times (1+z_j)]/max(T_{WRC4})$; $V_j$ is the comoving volume corresponding to the redshift bin $z_j$; $p(m_k)$ is the completeness limit for the WRC4-band; 
and where the luminosity function integral is performed within the luminosity limits corresponding to the magnitude bin $m_k$. 
Then, the total number of LAEs is $N_{\rm{LAE}}=\Sigma_j \Sigma_k N_{\rm{LAE}}(z_j,m_k)$. 
Based on this analyses, we find that the expected number of field LAEs in our field is $35\pm7$. In comparison, the observed number in our field is 65. The entire 0.317~deg$^2$ field is therefore overdense by $\approx 1.9\pm0.4$. Using the correct field density, we can update our earlier estimate, discussed in \S~\ref{galaxy_distribution}, of galaxy overdensity measured within a 8~Mpc radius circle as $\delta_\Sigma = 5.5-6.3$. 

\subsection{Present-Day Masses of the Protocluster Candidates}\label{cluster_mass}

If the galaxies trace the underlying mass distribution, what are the likely present-day descendants of these protocluster candidates? We can estimate the descendant masses using two approaches, one based on a comparison to numerical simulations, and the other based on a more heuristic approach. 

Recently, \citet{chiang13} used the Millennium Run simulations \citep{springel05} coupled with semi-analytic galaxy formation models to predict the expected (observable) overdensities associated with protoclusters of a given mass at $z>2$. \citet{chiang13} compute galaxy overdensities using a cubic tophat of volume (15~Mpc)$^3$. 
In order to compare our observations to their measurements, we measure the overdensity in a circular area 13~Mpc in diameter; since the line-of-sight depth of our survey is 25~Mpc, the total volume enclosed in our overdensity kernel is comparable (i.e., $\pi(13/2)^2 \times 25 \approx 15^3~{\rm Mpc^3}$). The galaxy overdensity is measured on the position of the three groups as $\delta_g=(N_{\rm{group}}/N_{\rm{field}})-1$ where $N_{\rm{field}}$ is 1.73, the mean number of LAEs within a 13~Mpc diameter circle based on the observed LAE surface density. The overdensity will be larger if we use the field density derived in the previous section. However, we continue to use the observed value in order to derive the most conservative estimates for the masses. The associated errors are estimated as a quadratic sum of Poissonian uncertainties ($1/\sqrt{N_{\rm{field}}}=76$\%) and cosmic variance of 52\% (see Appendix for calculation of cell (cosmic) variance). 
The S, NW and NE groups contain 9, 8, and 7 galaxies, respectively, and thus the overdensities in the S, NW and NE groups are  4.2$\pm$0.9, 3.6$\pm$0.9, 3.1$\pm$0.9, respectively.  \citet{chiang13} find that galaxy overdensities of $\delta_{\rm gal}=4.2$, 3.6, and 3.1 observed at $z\sim4$ will each evolve into $z=0$ galaxy clusters with total mass $M_{\rm tot,z=0} = (4-9)\times 10^{14}M_\odot$, $(2-6)\times 10^{14}M_\odot$, and $(2-4)\times 10^{14}M_\odot$, respectively (see their Figures 6, 10, and associated text). 

An alternative mass estimate can be made following the approach outlined by \citet{steidel98}. The mass associated with a galaxy overdensity that will eventually be bound and virialized by $z=0$ is given by:
\begin{equation}
\label{mass_estimate}
M_{z=0}=(1+\delta_m)\langle\rho\rangle V
\end{equation}
where $\rho$ is the average matter density of the universe ($=3H_0^2/8\pi G\Omega_0$),  $\delta_m$ is  matter overdensity, and $V$ is the volume enclosing the observed galaxy overdensity. With the adopted cosmology, Equation \ref{mass_estimate} is equivalent to $M_{z=0}=[3.67\times 10^{10}M_\sun]~ (1+\delta_m)~ [V/(1~{\rm Mpc})^3]$. The matter overdensity $\delta_m$ is related to the galaxy overdensity by $1+b\delta_m = C(1+\delta_{g})$, where $C$ is the correction factor that accounts for the effect of redshift-space distortions \citep{steidel98}.  In the case of spherical collapse, $C$ can be expressed as $C(\delta_m,z)=1+\Omega_0^{4/7}(z)[1-(1+\delta_m)^{1/3}]$. Within reasonable ranges of bias $b$ and $\delta_g$ ($b=1.8-4.0$, $2\leq \delta_g \leq 4$), the parameter $C$ ranges over $0.7-0.9$. We adopt a bias value for the LAE population of $b_{\rm LAE}\approx 2$ \citep[e.g.,][see also the Appendix of this paper]{gawiser07}. We estimate the volume $V$ to be the product of the effective transverse area containing the galaxy overdensities and the depth 25~Mpc.  The area is estimated from the iso-density contour map (Figure~\ref{structure_map}). 
For the NE and NW groups, the transverse area can be approximated as a 9~Mpc and 8~Mpc radius circle, containing 12 and 8 LAEs, respectively. The S group is more elongated as the density contour extends north towards additional 4 LAEs. Hence we approximate the area associated with the S group as a 10.3~Mpc$\times$25.6~Mpc rectangular region, containing 12 LAEs.

With the assumptions above, the total mass associated with each of the observed galaxy overdensities is $7\times10^{14}M_\odot$, $4\times10^{14}M_\odot$, and $8\times10^{14}M_\odot$ for the NE, NW, and S group, respectively. These values are higher than or close to the upper end of the mass estimate according to the \citet{chiang13} calibration. The main reason is because the Chiang et al. calibration assumed the galaxy bias of $b=2.35$ while we used the value $b=2.0$, directly measured from the data (see Appendix). Adopting their bias value returns the total masses closer to the Chiang et al. estimates; i.e., $6\times10^{14}M_\odot$, $4\times10^{14}M_\odot$, and $7\times10^{14}M_\odot$ for the NE, NW, and S group, respectively. Galaxy bias (hence, conversion from galaxy overdensity to matter overdensity) remains the largest uncertainty in our mass estimate. 

In summary, we estimate that the three galaxy overdensities have each $\delta_g\gtrsim 3$ measured within a $(15~\rm{Mpc})^3$ volume. \citet[][their Table~4]{chiang13} estimate that the regions with $\delta_g >3$  have $>80$\% probability to evolve into a galaxy cluster by $z=0$. Our mass estimates suggest that the NW group will evolve into a Fornax-like cluster ($M=(1-3)\times10^{14}M_\odot$), while the two more significant groups will evolve into Virgo-like clusters ($M=(3-10)\times10^{14}M_\odot$). The mass of the S group is nearly $10^{15}M_\odot$; if the overdensity extends further south, it is possible that its final mass may be even larger comparable to that of the Coma cluster (K.-S. Lee, et al., in preparation).  

\subsection{How Rare is this Region?}

We now consider the probability of finding {\it three} overdensities so close to one another. This is related to the expected number density of a single protocluster in our volume. According to \citet{chiang13}, the number of protoclusters identified to have $M>10^{14}M_\odot~h^{-1}$ at $z=0$ is 2,832 in a comoving box of 500~Mpc~$h^{-1}$ on a side. Hence, the expected number within our survey volume ($72\times72\times25~{\rm Mpc}^{3}$) is almost exactly one\footnote{We note that \citet{chiang13} used the simulations assuming the WMAP1 cosmology; i.e., $\Omega=0.25, \Omega_\Lambda=0.75$. We expect that the number of galaxy clusters in the WMAP7 cosmology, as assumed here, is slightly larger. } (i.e., $2832./(500./0.7)^3\times(72 \times 72 \times 25)=1.007$). An independent estimate from  the mass function of X-ray clusters gives a similar answer for the comoving space density \citep{reiprich02}. Given that the expected number is only one, what is the significance of an overdensity of 3 within a $\sim10^5~{\rm Mpc}^{3}$ volume? We estimate this by randomly sampling the Millennium Runs volume (assuming the WMAP7 cosmology) on scales of our survey box (Y.-K. Chiang, in private communication). Of $5\times10^{5}$ trials within the $3.6\times10^{8}~{\rm Mpc}^3$ volume, we find 2,800 cases in which three or more protoclusters are included in the box. Hence, the probability of finding three or more protoclusters {\it of any mass} within our volume is $\sim 5.7$\%. If we restrict the results to only groups where the total mass in protoclusters within the survey box is at least $10^{15}~M_\odot$, the probability decreases to 1.9\%. Similarly, the probability of finding three or more protoclusters within our survey volume where the most massive protocluster has a total mass of $5~(6,7)\times 10^{14}~M_\odot$ is 1.7 (1.2, 0.9)\%. In short, the chances of finding three overdensities matching our observational constraints within the survey volume are very low, $< 2$\%. We would only expect to make {\it one} such discovery if we carried out the LAE narrow-band survey to the same depth over $\gtrsim18$~deg$^2$! This structure may be one of the largest overdensities, both in size and density, observed thus far.

Although the protocluster candidates will likely virialize by $z=0$, the filamentary structures are unlikely to do so.  These filamentary regions will likely continue to evolve, perhaps becoming highly biased structures similar to local superclusters. The separation between the three overdensities is comparable to the size of a supercluster in its longest dimension \citep[see, e.g.,][]{einasto14}. Most extreme cases of such structures are known as  ``Great Walls'' in the local universe \citep{geller89, gott05}. To further elaborate on this hypothesis, we need to firmly establish the physical association  between different groups, which will require both larger areal coverage and kinematic information from spectroscopy. 

Finally, we checked the literature to find similarly massive structures containing multiple galaxy overdensities in the volume similar to our field. Most notably, \citet{chiang14} searched the 2~deg$^2$ COSMOS field  using the photometric redshift technique, and identified 36 protocluster candidates at $1.6<z<3.1$ (see their Table~1). While we do not find any structure containing three protoclusters, there are $\sim4$ cases where two candidates are within $<0.5$~deg and $\Delta z\leq 0.03$ from each other.  These structures lie at $z=2.07, 2.42, 2.74, 3.02$, and typically have $\delta_g  \sim 1.5-2.3$. The relatively low value for the observed overdensity is mainly due to the `dilution' caused by the photometric redshift uncertainty of typically $\Delta z \sim 0.1$ in identifying these structures (equivalent to $4-6\times$ larger  line-of-sight distance than our LAE survey). Based on their simulations covering $\sim24\times$ the COSMOS volume, \citet{chiang14} estimated that the probability of any of these candidate protoclusters to evolve into a Coma-like cluster ($M>10^{15}M_\odot$) is low, $ \lesssim 0.1$. 

Based on the number of protocluster pairs found by \citet{chiang14}, it is possible to independently estimate the likelihood of finding more than one protocluster within our survey volume. Over the 1.62~deg$^2$ area, they found four pairs with $\leq 0.5$~deg separation and  $\Delta z<0.03$. The comoving volume between $z=1.6$ and $z=3.1$ is 0.0196~Gpc$^3$. Therefore, the comoving number density of protocluster pairs is 126~Gpc$^{-3}$ ($=4/(0.0196\times 1.62)$). Since our survey volume is $72\times72\times25$~Mpc$^3=1.3\times 10^{-4}$~Gpc$^3$,  the expected number of a protocluster pair within our survey volume is 0.016. The estimated likelihood of $\sim2$\% is likely a conservative limit because the Bo\"otes structure contains three protoclusters; no triplet is found in the COSMOS protocluster sample. While more detailed comparison between the Bo\"otes structure and those found in the COSMOS field is not possible (Chiang et al. did not publish mass estimates for their candidates), it is assuring that the probability of finding multiple protoclusters in our survey volume is low at just a few percent based on both simulations and real data.  These considerations strongly suggest that the Bo\"otes structure represents one of the rarest overdensity regions with few known counterparts in the high-redshift universe. 

\section{Summary}\label{summary}
In this paper, we report the discovery of a large-scale structure consisting of three massive galaxy overdensities within a $72\times72\times25~{\rm Mpc}^3$ comoving volume at $z=3.78$. Deep imaging of the region taken with narrow- and broad-band filters reveal that the number of Lyman-alpha emitting galaxies (or LAEs) is $\sim90$\% higher than that expected in field environments. The LAEs in the field show a highly unusual angular distribution (Figure \ref{distribution} and \ref{structure_map}); three separate regions are observed with a surface overdensity, $\delta_\Sigma=(\Sigma-\bar{\Sigma})/\bar{\Sigma}$, of $\sim 3$ or higher within a 8~Mpc radius circle. These galaxy groups are separated about $40-60$~Mpc, and appear to be connected by filaments along which more LAEs reside. The distribution of LBGs in the same field resembles that of LAEs, even though the LBGs sample a factor of 20-25  larger line-of-sight distance than LAEs; this suggests that these overdensities are indeed  massive structures containing both LAEs and LBGs. The most significant overdensity lies at the extreme southern end of our image, suggesting that the true extent of this structure may extend beyond the current data. 

Among 65 LAEs, we find six  sources with high line-luminosity (${\rm WRC4}\leq 22.8$, implying $L_{\rm Ly\alpha}> 1.5\times10^{43}\ {\rm erg\ s^{-1}\ cm^{-2}}$). All but one of these high line-luminosity sources reside within or in the vicinity of one of the three galaxy overdensities. None of the 6 are detected at X-ray wavelengths, and the two that have been observed spectroscopically show no rest-frame UV signatures of AGN.  Hence, our data suggest that Ly$\alpha$ emission in these sources may be powered by massive star formation rather than by AGN. 

Recent galaxy simulations demonstrate that at the level of the observed overdensities, these structures will likely be virialized by $z=0$ and become galaxy clusters with masses $>10^{14}~M_\odot$ \citep{chiang13}. We estimate that one of the three galaxy groups will have the total mass of $(1-3)\times10^{14}M_\odot$ and the other two groups will grow in the range of  $(3-10)\times10^{14}M_\odot$. If the galaxy bias is lower than that assumed in the simulations, the final masses may be even larger. A direct estimate based on the sizes of the overdensities and measured galaxy bias in the region returns that the total masses enclosed within these regions may be as large as $\sim(3-7)\times10^{14}M_\odot$ for the two smaller groups, and $\sim (7-8)\times10^{14}M_\odot$ for the largest one. Given the relatively large distances between these groups ($\sim 40-60$~Mpc), it is unlikely that all three groups will merge into a single virialized structure by $z=0$. Nevertheless, it is rare to find three highly significant overdensities so close to one another. The Millennium simulation (volume $\sim 0.36$ billion Mpc$^3$) suggests that the probability of  chance alignment of three (or more) protoclusters within our survey volume ($\sim 10^5~{\rm Mpc}^3$) is $<$2\% (based on our conservative mass limits given in \S\ref{overdensity_estimate}). An independent estimate based on the existing samples of protocluster candidates at high redshift returned a consistently low likelihood. Hence, it appears that we have discovered one of the densest regions of the high-redshift universe with scales  well beyond $\gtrsim 40$~Mpc in its longest dimension, similar to the value observed for some of the local superclusters.  Further insight into the true extent and kinematic state of this structure, which will help us better assess its ultimate future, will require wider-field imaging data and spectroscopic followup. 

Identifying a statistical sample of very large structures and their constituents will provide a unique opportunity to study the early evolution of the universe. We will be able to study the early stages of galaxy formation in different local environments; from voids, to filaments, to the highest overdensity regions. Measurements of the space density of the protoclusters, and the galaxy distribution within these structures will directly test the theory of structure formation. Unfortunately, such large galaxy overdensities are extremely rare at high redshift and thus are unlikely to be found in large number from deep, small-area surveys. Targeted searches over much wider fields ($>50$~deg$^2$) will be key to identifying a statistical sample. Ongoing wide-field imaging surveys such as Dark Energy Survey, or future surveys with the Large Synoptic Survey Telescope, may provide a sound basis for selecting candidate protoclusters to be targeted.

\acknowledgments
KSL thanks Yi-Kuan Chiang and Roderik Overzier for providing the Millennium Runs data for comparison. This paper presents data obtained at the Kitt Peak National Observatory.  This research draws upon data provided by [Survey PI] as distributed by the NOAO Science Archive. NOAO is operated by the Association of Universities for Research in Astronomy (AURA) under cooperative agreement with the National Science Foundation.
We are grateful to the expert assistance of the staff of Kitt Peak National Observatory where the optical observations of the NDWFS Bo\"otes Field were obtained. The authors thank NOAO for supporting the NOAO Deep Wide-Field Survey. 
AD's research was supported in part by the National Optical Astronomy Observatory (NOAO) and by the Radcliffe Institute for Advanced Study and the Institute for Theory and Computation at Harvard University. NOAO is operated by the Association of Universities for Research in Astronomy (AURA), Inc. under a cooperative agreement with the National Science Foundation.

\appendix
\subsection{Estimating Galaxy Bias}
Galaxies are biased tracers of underlying matter distribution and estimates of the bias factor are necessary for converting observed galaxy overdensity $\delta_g$ to the underlying matter overdensity $\delta_m$. As a path to estimating the bias, we describe a procedure to measure ``cell variance''. Cell variance  can be easily predicted for any representative galaxy population (or dark matter) at any redshift and provides a powerful tool to characterize the large-scale environments in direct comparison with other known galaxy samples. 
We define the relative cell variance $\sigma^2_g$ as:
\begin{equation}
\sigma_g^2=\frac{\langle N^2 \rangle-\langle N \rangle^2}{\langle N \rangle^2}-\frac{1}{\langle N \rangle}
\end{equation}
where the first term is the total rms fluctuation in the number of galaxies found in a cell, and the second term accounts for the same quantity arising from the Poisson shot noise. Ideally, we want to measure cell variance in a spherical volume. Without redshift information of individual sources, it is not possible to pinpoint the LAE positions in the line-of-sight direction.  Because the evolution of clustering along the line-of-sight in our volume is negligible, the cell variance measured in a cylindrical volume should be as good as that measured in a sphere of the same volume. 

We measure cell variance by counting galaxies within a circle with radius 10~Mpc at random positions in our field. The volume sampled by the 10~Mpc circle is the same as that of a sphere of radius $8h^{-1}$~Mpc (=11.4~Mpc in the adopted cosmology). 
We correct the observed counts to account for the area lost to masked regions as $N_{\rm gal,corr} = N_{\rm gal}/(1 - f_{\rm loss})$, where $f_{\rm loss}$ is the fraction of the total area that is masked out; we exclude measurements where the loss is more than 60\% as unreliable. When averaged over the corrected number counts made over 1000 random trials, we recover the field-averaged surface density within 3\%. 

We measure $\sigma_g=0.44\pm0.05$ for our sample. The power spectrum normalization for the adopted cosmology is $\sigma_8=0.8$, i.e., the relative fluctuation of matter at $z=0$ within a $8~h^{-1}$ radius sphere. In the linear regime, density perturbations grow by a factor of 3.64 from $z=3.785$ to $z=0$. (see Equation~19 of \citet{carroll92}); i.e., matter fluctuation  is $\sigma_{8} (z=3.785)=0.22$. Hence, we estimate that galaxy bias for our LAEs is $\sim2.0\pm0.2$ ($=[0.44\pm0.05]/0.22$). Our value is consistent with the bias estimated from clustering measurements of field LAEs \citep{gawiser07}, but is slightly lower than the predictions from galaxy simulations \citep[e.g.,][]{chiang13} derived from the Millennium runs, $b\sim 2.35$, for all galaxies satisfying the condition ${\rm SFR} > 1M_\odot~{\rm yr}^{-1}$.

\end{document}